\documentclass[preprint,showpacs,preprintnumbers,amsmath,amssymb]{revtex4}

\usepackage{amsmath}
\usepackage[dvips]{graphicx}
\usepackage{subfigure}
\usepackage{bm}
\usepackage{graphicx}
\usepackage{dcolumn}

\begin{document}

\preprint{APS/123-QED}

\title{A coarse grained model of polymer networks \\focusing on the intermediate length scales} 

\author{Takashi Shibata}
\altaffiliation{email:t-shibata@cmpt.phys.tohoku.ac.jp} 
\author{Hidemitsu  Furukawa$^{+}$}
\author{Toshihiro Kawakatsu}%
\affiliation{Physics Department, Tohoku University , Sendai 980-8578, Japan,\\and Hokkaido University, Sapporo 001-0021, Japan$^{+}$}
\date{2008/3/22}

\begin{abstract}
We propose a coarse-grained model for polymer chains and polymer networks based on the meso-scale dynamics. 
The model takes the internal degrees of freedom of the constituent polymer chains into account using memory functions and colored noises.
We apply our model to dilute polymer solutions and polymer networks.
A numerical simulation on a dilute polymer solution demonstrates the validity of the assumptions on the dynamics of our model. 
By applying this model to polymer networks, we find a transition in the dynamical behavior from an isolated chain state to a network state.
  
\end{abstract}

\pacs{Valid PACS appear here}
\maketitle


A wide class of systems possesses hierarchical structures over large length scales.
Typical examples are critical liquids and soft matter such as polymers, surfactants, colloidal suspensions, and polymer gels. 
Due to the coexistence of the different length scales of the internal freedom degree, soft matter shows various anomorous and interesting phenomena including shear induced phase separation of polymer solutions \cite{test1}\cite{test2}, and viscoelastic phase separations of  entangled polymers \cite{test3}.
In such point of view, the system of polymer gels is a somewhat interesting target, because it has widely distributed length scales, i.e. the size  of monomers, the size of  networks, and the size of the whose elastic body.
Due to such hierarchical structures, many interesting phenomena such as swelling behavior coupled with inhomogeneity \cite{batterfly}\cite{Furukawa1} and anomalous relaxation process in the dynamics of networks\cite{critical slowing down}\cite{topologycal relax} occur.   
The hierarchical structures of polymer gels are also utilized to develop many innovative materials\cite{topological}\cite{cray}\cite{double network}.
  
To realize these phenomena and to design these materials, it is important to model meso-scale structures and to bridge between these mesoscale structures and those on the larger and smaller scales.
There have been many models for the polymer network systems on different length scales\cite{Polymer Science}.
However, due to the complexity of the polymer networks, these models are far from realistic especially on the intermediate length and time scales, which are important in understanding the experimental results and in materials designing.

In this paper, we propose a coarse-grained model of polymer networks based on the meso-scale structures and dynamics.
We perform numerical simulations of this model to verify the validity of our assumptions and modeling.

\begin{figure}[b]
\begin{center}
\includegraphics [width=6.5cm,clip]{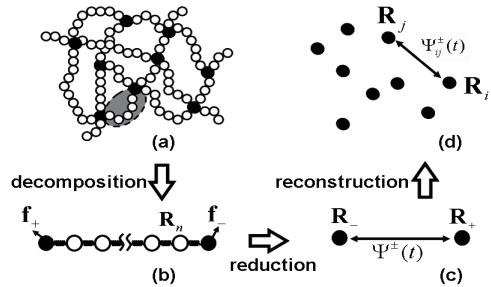}
\end{center}
\caption{Schematic diagram of the reduction process of the degrees of freedom of polymer network. 
(a) A polymer network before the reduction.
(b) A tagged single polymer chain in the polymer network. 
(c) The polymer chain described by memory functions $\Psi^{\pm}$. 
(d) Polymer network after the reduction.}
\label{fig:model_picture}
\end{figure}

Let us discuss our coarse-grained model for polymer networks. 
To reduce the degrees of freedom, we derive a set of dynamical equations for the polymer networks which are described in terms of the degrees of freedom of the crosslinkers.
In Fig.\ref{fig:model_picture}, we show a schematic illustration of our reduction process of the degrees of freedom.
Here we model the polymer network as a set of linear polymer chains connected by crosslinkers as is shown in Fig.\ref{fig:model_picture}(a).
We can describe this network topology using so-called adjacency matrix.
To obtain explicit expressions for the memory functions and the colored noises based on  the meso-scale polymer network model, 
we decompose polymer network into a set of linear polymer chains which are described by the Zimm model
 for the dilute polymer solution under hydrodynamic interaction(Fig.\ref{fig:model_picture}(b))\cite{doi-ed}.
By a reduction of the degree of freedom of the dynamic equation for the linear polymer chain, 
we gain the memory functions and colored noises for the polymer chain(Fig.\ref{fig:model_picture}(c)).   
Then, we reconstruct the polymer network by connecting these chains using the adjacency matrix.
With this procedure, we can express the motion of the crosslinkers without using information on the motion of the monomers 
(Fig.\ref{fig:model_picture}(d)). After this reduction, the degrees of freedom of the monomers are reduced to the memory functions and colored noises\cite{Moriformula}.

\begin{figure}[t]
\begin{center}
\includegraphics [width=8cm,clip]{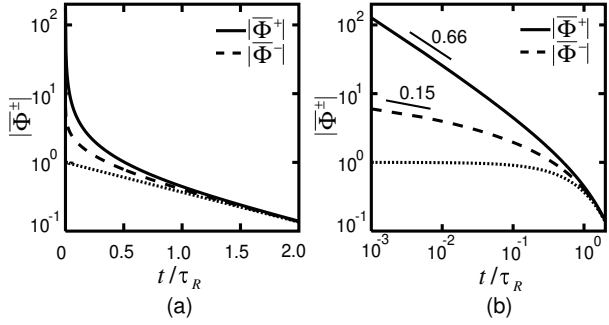}
\end{center}
\caption{Time dependence of the normalized memory kernels $\bar{\Phi}^{\pm}$ which are defined by $\bar{\Phi}_{\pm}=\Phi_{\pm}/A_{\pm}t_0^{\alpha_{\pm}}$   
on (a) semi-log plot and on (b) log-log plot. 
The top (solid) curve is $\bar{\Phi}^{+}$, the middle (broken) curve is $\bar{\Phi}^{-}$, 
and the bottom (dotted) curve shows the single exponential function for comparison.}
\label{fig:phi}
\end{figure}

Now, we discuss the dynamics of an isolated linear polymer chain, which is composed of $N$ monomers.
Let $\mathbf{R}(n,t),(n=1,...N-1)$ be the position of $i$-th monomer at time $t$, 
and define $\mathbf{R}_+(t)=\mathbf{R}(0,t)$, $\mathbf{R}_-(t)=\mathbf{R}(N,t)$, 
$\mathbf{v}_\pm=d\mathbf{R}_\pm/dt$, and $\mathbf{f}_{\pm}$
as the positions and the velocities of crosslinkers at the chain ends and the forces  acting on  them.
In this dilute polymer case, the dynamics of monomers are well described by the Zimm model\cite{doi-ed}.
In the Zimm model, the equation of motion of individual monomer is expressed as 
\begin{align}
\label{Zimm model about linear polymer}
&\frac{\partial \mathbf{R}_n}{\partial t} =k\int_0^N dm \mspace{5mu} h(n-m) \frac{\partial ^2 \mathbf{R}_m}{\partial m^2} 
                             + \bm{\mathbf{\eta}}_n + \mathbf{f}_n \\
\label{Zimm model fluctuation dispation theorem}
&\langle \bm{\mathbf{\eta}}_n(t) \bm{\mathbf{\eta}}^T_m(t') \rangle = 2k_BT h(n-m)\bm{I}\delta(t-t')
\end{align}
where $h(n-m)$ and $\bm{\mathbf{\eta}}_n$ are  
mobility of the  monomer and the noise acting on the $n$-th monomer and $\mathbf{f}_n$ is defined as $\mathbf{f}_n=\mathbf{f}_+N\delta (n) + \mathbf{f}_-N\delta (n-N)$. 
The mobility of the monomers is related to the noise $\bm{\mathbf{\eta}}_n$ 
by the fluctuation dissipation theorem 
eq.\eqref{Zimm model fluctuation dispation theorem}, where $\bm{I}$ is the unit tensor.
In order to solve eq.\eqref{Zimm model about linear polymer} formally, we introduce the Fourier series expansion of any vector variable $\mathbf{Q}_n$ such as $\bm{\mathbf{R}}_n$  as
$\mathbf{Q}_n=\tilde{\mathbf{Q}}_0+2 \sum_{p=1}^N \tilde{\mathbf{Q}}_p\cos (\frac{p\pi n}{N})$ 
 together with the  expressions for the Fourier coefficients $\tilde{\mathbf{Q}}_0=\frac{1}{N}\int_0^N dn \mathbf{Q}_n$ and $\tilde{\mathbf{Q}}_p=\frac{1}{N}\int_0^N dn \mathbf{Q}_n\cos (\frac{p\pi n}{N})$.
Here we neglected the sine modes in the Fourier series, because we focus on the dynamics of the end points of the polymer chain which do not excite the cosine modes. 
Then the equation of motion eq\eqref{Zimm model about linear polymer} is rewritten as
\begin{align}
\label{Zimm model about linear polymer to mode}
\frac{\partial}{\partial t}\tilde{\mathbf{X}}_p = \zeta_p^{-1}(-k_p \tilde{\mathbf{X}}_p + \tilde{\mathbf{f}_p}) + \tilde{\bm{\mathbf{\eta}}}_p,
\end{align}	
where $k_{p}=6\pi^2k_BT(Nb^{2})^{-1}p^2$ and $\zeta_{p}=(12\pi^3 N b^2 p)^{1/2}\eta_{s}$.	
Here, the longest relaxation time $\tau_R=\zeta_1/k_1$ is called Rouse time.
If we only focus on the dynamics of crosslinkers, 
the independent variables  we need are $\mathbf{R}_+(t)=\mathbf{R}(0,t)$ and $\mathbf{R}_-(t)=\mathbf{R}(N,t)$.
Since eq.\eqref{Zimm model about linear polymer to mode} is linear equation,  
 we can  directly integrate eq.\eqref{Zimm model about linear polymer to mode} over $t$.  
Then, we can sum up Fourier coefficients $\tilde{\mathbf{X}}_p$ to derive $\mathbf{R}_{\pm}$. 
As a result, the velocities of crosslinkers $\mathbf{v}_{\pm}=d\mathbf{R}_{\pm}/dt$ are given by
\begin{align}
\label{R_pm}
\mathbf{v}_{\pm}=\int ds\{g_+(t-s)\mathbf{f}_{\pm}(s)+g_-(t-s)\mathbf{f}_{\mp}(s)\}+\tilde{\bm{\mathbf{\xi}}}_{\pm} ,
\end{align}	
where $g_{\pm}(t)$ and $\xi_{\pm}(t)$ are the memory functions and colored  noises that are defined by
\begin{align}
\label{Green function Gpm}
g_{\pm}(t) &= \frac{1}{\zeta_0}+\sum_{p=1}^N \frac{2}{\zeta_p} 
\bigg[1 - \frac{({\pm}1)^p p^{3/2}}{\tau_R} \exp \big({-\frac{tp^{3/2}}{\tau_R}}\big) \bigg], \\
&\langle \tilde{\bm{\xi}}_{\pm}(t) \tilde{\bm{\xi}}^T_{\pm}(t') \rangle = k_BT g_{+}(t-t'),  \\
&\langle \tilde{\bm{\xi}}_{\pm}(t) \tilde{\bm{\xi}}^T_{\mp}(t') \rangle = k_BT g_{-}(t-t') .
\end{align}		
Using eq.\eqref{R_pm}, the force acting on the crosslinkers which is caused by the motion of internal degrees of freedom of the polymer chain can be described  as
\begin{align}
\label{f_pm_final}
\mspace{-11mu}\mathbf{f}_{\pm}
   =\int ds \{ \Phi^+(t-s) \mathbf{v}_{\pm}(s) +\Phi^-(t-s) \mathbf{v}_{\mp}(s) \}+ \bm{\xi}_{\pm}.
\end{align}	
The memory functions $\Phi^{\pm}$ are given by 
\begin{align}
\label{g and Phi}
\Phi^{\pm}(\omega)=\pm g_{\pm}(\omega) / (g_+^2(\omega)-g_-^2(\omega)),
\end{align}
where $\Phi^{\pm}(t)$ are calculated from eqs.\eqref{Green function Gpm} and \eqref{g and Phi} numerically.
In Fig.\ref{fig:phi}, we show the result of the numerical evaluation of $\Phi^{\pm}(t)$.
We confirm that the memory functions $\Phi^{\pm}(t)$ can be very well approximated by
\begin{align}
\label{Phi_pm_2}
\Phi^{\pm}(t;\tau_R) =\mp A_{\pm} (t/t_0)^{-\alpha_{\pm}}\exp  (-t/\tau_R), 
\end{align}	
where $t_0$ are the characteristic time scales of the microscopic dynamics, 
and the numerical values of exponent $\alpha_{\pm}$ are found to be $\alpha_{+} = 0.66$ and $\alpha_{-} = 0.15$, respectively.

Now, we derive a coarse-grained model for polymer networks based on the Langevin equation for a single chain
eqs.\eqref{R_pm} \eqref{f_pm_final} and \eqref{Phi_pm_2}.
Let $\mathbf{R}_i(t)$ and $\mathbf{v}_i(t)$ be the position and the velocity of $i$-th crooslinker at time $t$,
 and $\mathbf{F}_i(t)$ be the force acting on it.
We define $\mathbf{f}_{ij}$ the force acting on the $i$-th crosslinker imposed by the $j$-th crosslink through the chain connecting them.
Since the system we discuss is composed of the linear polymer chains which obey the linear integral equation 
eq.\eqref{f_pm_final}, we can derive the force $\mathbf{F}_{i}$
as $\mathbf{F}_i=\sum_{j=0}^N A_{ij} \mathbf{f}_{ij}$ using linear superposition with the adjacency matrix $A_{ij}$ for the polymer network. 
If $i$ and $j$ is connected, $(i,j)$ element of the adjacency matrix  is defined as $A_{ij}=1$. On the other hand, $A_{ij}=0$, if $i$ and $j$ are unconnected.
The equations of motion of the crosslinkers are given by 
\begin{align}
\label{general langevin 6}
\frac{d}{d t}\mathbf{R}_{i} &= \mathbf{v}_{i}, \\
\label{general langevin 7} 
m\frac {d\mathbf{v}_i}{dt}&=\sum_j \int_{-\infty}^{t} \mspace{-15mu} ds \mathcal{M}_{ij}(t-s) \mathbf{v}_{j}(s) 
+k\sum_j \Omega_{ij}\mathbf{R}_{j}+\bm{\xi}_{i},\\
\label{momory_karnel}
\mathcal{M}_{ij} &=    
 \bigg(\sum_k A_{ik} \Phi^+_{ik}-\zeta\bigg) \delta_{ij}+A_{ij}\Phi^-_{ij}.
\end{align}
The memory functions $\Phi_{ij}^{\pm}$ are defined as 
$\Phi_{ij}^{\pm}= {\mp}A_{\pm}(t/t_0)^{-\alpha_{\pm}}\exp  (-t/\tau_{ij}) $,
and $\tau_{ij}$ is the  maximum relaxation time of the chain between 
$i$-th and $j$-th crosslinkers.
$\zeta$ and $k$ are friction coefficient of a crosslinker and spring constant of a chain. 
Because the viscosity term in eq.\eqref{general langevin 7} is in general much larger than the inertia term in polymer solution,  we can neglect the latter term.
Finally, we discretize eqs.\eqref{general langevin 6} and \eqref{general langevin 7} so that we can integrate them numerically by Euler method as
\begin{align}
&\label{discretize form 1}
\mathbf{R}_{i}(t+\Delta t) = \mathbf{v}_{i}(t) \Delta t + \mathbf{R}_{i}(t) + O(\Delta t^2), \\ 
\label{discretize form 2} 
&\mathbf{v}_i(t)=\sum_{j,k}\bigg[\int_0^{\Delta t} \mspace{-15mu}  ds\bm{\mathcal{M}}(s)\bigg]^{-1}_{ik} \notag \\
	&\mspace{15mu} \times\bigg[\sum_{n=1}^{\infty}\Delta t\mathcal{M}_{kj}(n\Delta t)\mathbf{v}_j(t-n\Delta t)+k\Omega_{kj}\mathbf{R}_j(t)
+\bm{\mathbf{\xi}}_k(t)\bigg], \\ 
\label{discretize form 3}
&\langle  \bm{\mathbf{\xi}}_i(t) \bm{\mathbf{\xi}}^T_j(t+n\Delta t) \rangle  \notag \\
&\mspace{15mu}= 2k_B T \bm{I} \bigg\{ \bigg[\int_0^{\Delta t} \mspace{-15mu} ds\bm{\mathcal{M}}(s)\bigg]^{-1}_{ij}  \mspace{-15mu}  \delta_{0,\Delta t} + \mathcal{M}_{ij}^{-1}(n\Delta t) \bigg\},
\end{align}
where the memory kernel $\mathcal{M}_{ij}(t)$ is related to the colored noise $\bm{\mathbf{\xi}}_i(t)$ 
by the fluctuation dissipation theorem eq.\eqref{discretize form 3}.


We calculate the mean-square displacement $C(t)$, intermediate scattering function of the position  of the crosslinkers $g_c(\mathbf{q},t)$, 
and the total scattering from both the crosslinkers  and the  monomers $g(\mathbf{q},t)$ 
by molecular dynamics (MD) simulations using eqs.\eqref{discretize form 1} and \eqref{discretize form 2}.
Here, $C(t)$, $g_c(\mathbf{q},t)$ and $g(\mathbf{q},t)$ are defined by
\begin{align}
\label{C_t_define}
&C(t)=\frac{1}{N}\sum_{i} \langle |\mathbf{R}_i(t)-\mathbf{R}_i(0)|^2 \rangle, \\
\label{g_c_t_define}
&g_c(\mathbf{q},t)=\sum_{i,j} \langle\exp\big[i\mathbf{q}\cdot(\mathbf{R}_i(t)-\mathbf{R}_j(0))\big]\rangle, \\
\label{g_t_define}
&g(\mathbf{q},t)=\sum_{i,j,k,l}\langle S_{ij}(\mathbf{q})S^{*}_{kl}(\mathbf{q})\rangle\langle\exp\big[i\mathbf{q}\cdot(\mathbf{X}_{ij}(t)-\mathbf{X}_{kl}(0))\big]\rangle,
\end{align}
where $\mathbf{R}_i$ and $\mathbf{X}_{ij}=(\mathbf{R}_i+\mathbf{R}_j)/2$ are the position of the $i$-th crosslinker andthe position of center of mass of  the chain between crosslinkers $i$ and $j$.
Here, $S_{ij}(\mathbf{q})$ expresses the scattering from the chain connecting 
$i$-th and $j$-th crosslinkers, and defined as 
$S_{ij}(\mathbf{q})=\sum_{m} \exp(i\mathbf{q} \cdot \mathbf{r}_m)$, with $\mathbf{r}_m$ being the position of $m$-th monomer that belongs to the chain between $i$-th and $j$-th crosslinkers.
In eq.\eqref{g_t_define}, we neglect  correlation between the  internal degrees of freedom of each chain and that of its  center of mass.
Then, $\langle S_{ij}(\mathbf{q})S^{*}_{kl}(\mathbf{q})\rangle=\langle |S_{ij}(\mathbf{q})|^2 \rangle\delta_{ik}\delta_{jl} $ is approximated as Debye function, 
because the relaxation of $\langle S_{ij}(\mathbf{q})S^{*}_{kl}(\mathbf{q})\rangle$ is much faster than  that of the correlation between 
 the positions of centers of mass.  

To confirm the validity of our model, we first calculate $C(t)$  for a single linear polymer chain, whose ends are labeled as $1$ and $2$.
In this case, we set $\Delta t =0.001$, $A_{\pm}t_0^{\alpha_{\pm}}=1.1$, $\zeta=0.25$, $k=0.075$, $k_B T=1.0$ and $\tau_{12}=1.0$,
where we measure length and time in the units of $\tilde{x}=R_g$ and $\tilde{t}=\tau_{12}=\tau_R$,
 where $R_g$ is the radius of gyration of the polymer chain. 
For this linear polymer, the adjacency matrix and the interaction matrix are given by

$	
    \bm{A}= 
	\left(
	\begin{array}{@{\,}cc@{\,}}
	  0 & 1 \\
	  1 & 0 \\
	\end{array}
    \right)
$
\hspace{10mm}
$	\bm{\Omega}
    = 
	\left(
	\begin{array}{@{\,}cc@{\,}}
	  -1 & 1 \\
	  1 & -1 \\
	\end{array}
    \right).
$

\begin{figure}[t]
\begin{center}
\includegraphics [width=8.0cm,clip]{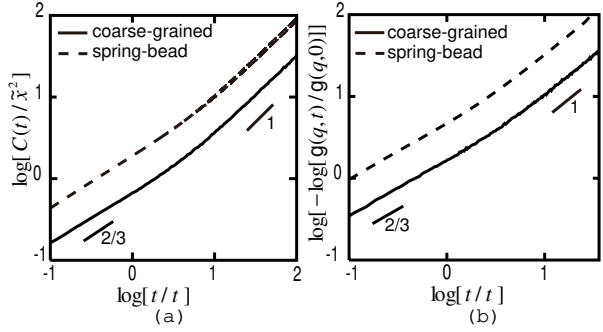}
\end{center}
\caption{(a)Auto-correlation function of the position of the one  end point of a linear polymer chain. 
(b)Dynamic structure factor of a strand composed of a linear polymer chain.
The solid line is result of coarse-grained model, 
and dotted line shows spring-beads model's one, where the dotted line is scrolled up for comparison.}
\label{fig:C_t}
\end{figure}

\noindent The simulation results are shown in Fig.\ref{fig:C_t}(a) along with the data of the non coarse-grained bead-spring model that will be described later.
It is evident that in the short time regime ($t\ll\tilde{t}$) the auto-correlation function $C(t)$ behaves  as $C(t)\sim t^{2/3}$, 
while in the long time regime ($t\gg\tilde{t}$) this function can be approximated as $C(t) \sim t$.
These results are consistent with the experimental study\cite{DNA} and the scaling argument\cite{preparation}.
Therefore we confirm the validity of our model for a single polymer chain.
The intermediate scattering function eq.\eqref{g_t_define} for a dilute polymer solution is also calculated 
and is shown in Fig.\ref{fig:C_t}(b). Here, we set $\mathbf{q}$  as  
$|\mathbf{q}|R_g=2 \pi$  . 
In the short time regime, the relaxation of $g(\mathbf{q},t)$ is well described as the  stretched exponential function
 $g(\mathbf{q},t) \sim g(\mathbf{q},0) \exp (-\Gamma_1 t^{2/3})$.
On the other hand, in the long time regime, we can fit  $g(\mathbf{q},t)$ as $g(\mathbf{q},t)\sim g(\mathbf{q},0) \exp (-\Gamma_2 t)$.
From this numerical result, it is clear that the intermediate scattering function  $g(\mathbf{q},t)$ can be described only using 
the degrees of freedom of the crosslinkers.
In this simulation, we also compare our model to the spring-bead model (non coarse-grained model).
In Fig.\ref{fig:C_t}, it is shown that our model  correctly reproduces the non coarse-grained one, 
in spite of its degrees of freedom is smaller than the non coarse-graind one\cite{preparation}.
\begin{figure}[t]
\begin{center}
\includegraphics [width=8cm,clip]{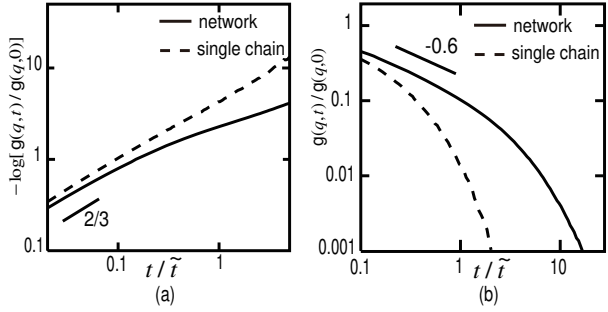}
\end{center}
\caption{ Time dependence of intermediate scattering function of the position  of the crosslinkers $g_c(t)$ 
in (a) short time regime and in (b) intermediate time regime.
The solid line is the simulation result of polymer network, and the dotted line shows that of single polymer chain for comparison.
}
\label{fig:network}
\end{figure}
 
Next, we apply this model to a polymer network to understand the hierarchical structure.
In the simulation, we use the same parameters as those used in Fig.\ref{fig:C_t}, and  we set the number of crosslinkers as $N$ = 1000.
To describe connectivity of the polymer network systems, we prepare 
a set of $N$ linear  polymers, and  connect any pair of the end points randomly with a probability $p$.
Here, we set the probability $p=0.0005$, which corresponds to the percolation point of this random network system \cite{preparation}. 
The interaction matrix $\Omega_{ij}$ is given by $\Omega_{ij}=-(\sum_k A_{ik}) \delta_{ij}+A_{ij}$.
In Fig.\ref{fig:network}, we show the result of simulations on $g_c(\mathbf{q},t)$ with $|\mathbf{q}|R_g=2 \pi$.
This numerical result shows that in short time regime ($t/\tilde{t}<0.1$), the correlation function of the network is the same as the one for a single chain.  
On the other hand, in the intermediate time regime ($0.1<t/\tilde{t}<1$), 
the relaxation of this function is much slower than that of a single chain,  
and this is roughly approximated by a power low function as $t^{-0.6}$.
This power law relaxation process can be explained by a simple scaling approach using the cluster distribution function $P(n) \sim n^{\alpha}$ and the maximum relaxation time $\tau_{\textrm{max}}(n)\sim n^{\beta}$ at a percolation threshold.
Since this percolated system can be classified as Bathe grid system, 
 the cluster distribution function $P(n)$ and the cluster radius $R_s$ 
is described in general as $P(n)\sim n^{-2.5}$ and $R_s\sim n^{0.25}$.
The relation between $\tau_{\textrm{max}}$, $R_s$ and 
number of monomers form end to end of the cluster $N$ are $\tau_{\textrm{max}}\sim N^2 \sim R_s^{4}$, and hence $\beta=1$. In the present situation, because of finite size effect and hydrodynamic interaction between crosslinkers, the exponent $\beta$ is slightly smaller than this expected value and is estimated as $\beta \sim 5/6$.
Then, in the intermediate time regime $g_c(t)$ is described as
$g_c(t)\sim \int dn \mspace{5mu} nP(n)e^{-t/\tau_{\textrm{max}}(n)} \sim t^{(\alpha+2)/\beta}=t^{-0.6}$,which reproduces the power low's behavior shown in Fig.\ref{fig:network}(b).
Thus, we conclude that the origin of the power law relaxation in the intermediate time scale (Fig.\ref{fig:network}(b)) is the motion of percolated clusters.
Similar phenomena were observed experimentally under the gelatin process\cite{critical slowing down}.

In conclusion, we have proposed a coarse-grained model for polymer networks and showed its numerical results.  
The main results of the present work are summarized below.
(i) We have derived a set of dynamic equations of polymer networks only using the degrees of freedom of the crosslinkers, and obtained the expressions of the quite general memory kernels and the random noises.
(ii) The auto-correlation function $C(t)$ of the crosslinkers at the both sides of a single polymer chain and their intermediate scattering function $g(\mathbf{q},t)$ in dilute polymer solution were calculated numerically.
 The simulation results were consistent with the experimental and theoretical results \cite{DNA}.
(iii) We applied this model to a polymer network, and succeeded in smoothly connecting the dynamics of crosslinkers between \textsf{"}the internal motion of a single chain\textsf{"} and \textsf{"}the motion of percolated clusters\textsf{"}.
 
We thank Yoshinori Hayakawa, Nariya Uchida, Katsuhiko Sato, Hiroto Ogawa, and Kenji Ohira for variable discussions and useful comments.
This work is supported by the 21 Century COE Program of Tohoku Univ. and Grant in Aid for Priority Area Research "Soft Matter Physics" (No.463) from MEXT of Japan.

\end{document}